# Reduction of Blocking Artifacts
# In
# JPEG Compressed Image

*Sukhpal Singh*
*B.Tech. (C.S.E.)*
*Computer Science and Engineering Department*
*Guru Nanak Dev Engineering College, Ludhiana, Punjab, India*

**ABSTRACT**

In JPEG (DCT based) compresses image data by representing the original image with a small number of transform coefficients. It exploits the fact that for typical images a large amount of signal energy is concentrated in a small number of coefficients. The goal of DCT transform coding is to minimize the number of retained transform coefficients while keeping distortion at an acceptable level. In JPEG, it is done in 8X8 non overlapping blocks. It divides an image into blocks of equal size and processes each block independently. Block processing allows the coder to adapt to the local image statistics, exploit the correlation present among neighboring image pixels, and to reduce computational and storage requirements. One of the most degradation of the block transform coding is the "blocking artifact". These artifacts appear as a regular pattern of visible block boundaries. This degradation is a direct result of the coarse quantization of the coefficients and the independent processing of the blocks which does not take into account the existing correlations among adjacent block pixels. In this paper attempt is being made to reduce the blocking artifact introduced by the Block DCT Transform in JPEG.

**INTRODUCTION**

As the usage of computers continue to grow, so too does our need for efficient ways for storing large amounts of data (images). For example, someone with a web page or online catalog – that uses dozens or perhaps hundreds of images-will more likely need to use some form of image compression to store those images. This is because the amount of space required for storing unadulterated images can be prohibitively large in terms of cost. Methods for image compression are lossless and lossy image compression. The JPEG is a widely used form of lossy image compression standard that centers on the Discrete Cosine Transform (DCT). The DCT works by separating images into parts of differing frequencies. During a step called quantization, where part of compression actually occurs, the less important frequencies are discarded. Then, only the important frequencies remain and are used to retrieve the image in the decompression process. The reconstructed images contain some distortions. At low bit-rate or quality, the distortion called blocking artifact is unacceptable. This dissertation work deals with reducing the extent of blocking artifacts in order to enhance the both subjective as well as objective quality of the decompressed image.

**Motivation**

JPEG defines a "baseline" lossy algorithm, plus optional extensions for progressive and hierarchical coding. Most currently available JPEG hardware and software handles the baseline mode. It contains a rich set of capabilities that make it suitable for a wide range of applications involving image compression. JPEG requires little buffering and can be efficiently implemented to provide the required processing speed. Also, implementations can trade off speed against image quality by choosing more accurate or faster-but-less-accurate approximations to the DCT. Best Known lossless compression methods can compress data about 2:1 on average. Baseline JPEG (**color images at 24 bpp**) can typically achieve 10:1 to 20:1 compression without visible loss and 30:1 to 50:1 compression visible with small to moderate defects. **For Gray Images (at 8 bpp)**, the threshold for visible loss is often around 5:1 compression. The baseline JPEG coder is preferable over other standards because of its low complexity, efficient utilization of memory and reasonable coding efficiency. Being owner of such features, JPEG is a leading image format used on the Internet and at home for storing high-quality photographic images, as well as the image format of choice for storing images taken by digital cameras. But JPEG suffers from a drawback- blocking artifacts, which are unacceptable in the image at low bit-rates.Although more efficient compression schemes do exist, but JPEG is being used for a long period of time that it has spread its artifacts over all the digital images. The need for a blocking artifact removal technique is therefore a motive that constantly drives new ideas and implementations in this field.Considering the wide spread acceptance of JPEG standard, this dissertation suggested a post processing algorithm that does not make any amendments into the existing standard, and reduces the extent of blocking artifacts. That is, the work is being done to improve the quality of the image.

**Image Compression**

As the beginning of the third millennium approaches, the status of the human civilization is best characterized by the term "Information Age". Information, despite its physical non-existence can dramatically change human lives. Information is often stored and transmitted as digital data. However, the same information can be described by different datasets. The shorter the data description, usually the better, since people are interested in the information and not in the data. Compression is the process of transforming the data description into a more concise and condensed form. Thus improves the storage efficiency, communication speed, and security. Compressing an image is significantly different than compressing raw binary data. Of course, general purpose compression programs can be used to compress images, but the result is less than optimal. This is because images have certain statistical properties which can be exploited by encoders specifically designed for them.

**DATA =
REDUNDANT DATA +
INFORMATION**

**Figure. 1: Relationship between data and information**

**Motivation behind image compression**

A common characteristic of most images is that the neighboring pixels are correlated and therefore contain redundant information. The foremost task then is to find less correlated representation of the image. In general, three types of redundancy can be identified:

a. Coding Redundancy

If the gray levels of an image are coded in a way that uses more code symbols than absolutely necessary to represent each gray level, the resulting image is said to have coding redundancy.

b. Interpixel Redundancy

This redundancy is directly related to the interpixel correlations within an image. Because the value of any given pixel can be reasonably predicted from the value of its neighbors, the information carried by individual pixels is relatively small. Much of the visual contribution of a single pixel to an image is redundant; it could have been guessed on the basis of the values of its neighbors.

c. Psychovisual Redundancy

This redundancy is fundamentally different from other redundancies. It is associated with real or quantifiable visual information. Its elimination is possible only because the information itself is not essential for normal visual processing. Since the elimination of psycho visually redundant data results in a loss of quantitative information, it is commonly referred to as quantization.

**Image Compression Model**

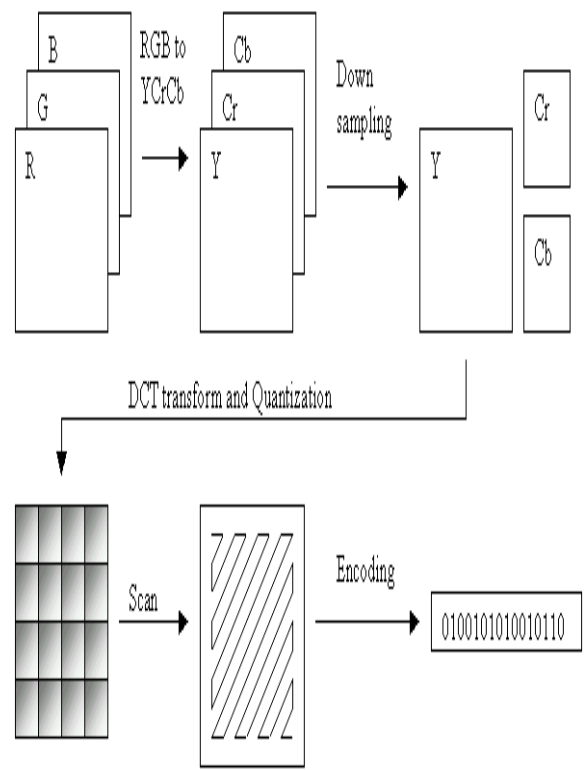

**Fig. 2 Image compression Model**

As the above figure 2 shows, a compression system consists of two distinct structural blocks: an encoder and a decoder. An input image f(x,y) is fed into the encoder, which creates a set of symbols from the input data. After transmission over the channel, the encoded representation is fed to the decoder, where the reconstructed output image f'(x,y) is generated. In general, f'(x,y) may or may not be the exact replica of f(x,y).The encoder is made up of a source encoder, which removes input redundancies, and a channel encoder, which increases the noise immunity of the source encoder's output same is in the case of decoder, but functions in reverse direction.

**Compression Techniques**

There are two different ways to compress images-lossless and lossy compression.

**Lossless Image Compression**: A lossless technique means that the restored data file is *identical* to the original. This type of compression technique is used where the loss of information is unacceptable. Here, subjective as well as objective qualities are given importance. In a nutshell, decompressed image is exactly same as the original image.

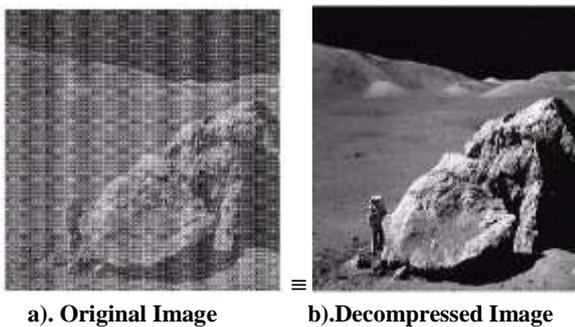

a). Original Image ≡ b).Decompressed Image
**Fig. 3 Relationship between input and output of Lossless Compression**

**Lossy Image Compression:** It is based on the concept that all real world measurements inherently contain a certain amount of *noise*. If the changes made to these images, resemble a small amount of additional noise, no harm is done. Compression techniques that allow this type of degradation are called **lossy**. This distinction is important because lossy techniques are much more effective at compression than lossless methods. The higher the compression ratio, the more noise added to the data.

In a nutshell, decompressed image is as close to the original as we wish.

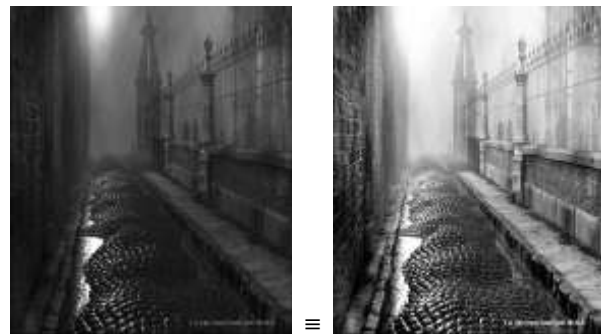

a).Original Image ≡ b).Decompressed Image
**Fig. 4 Relationship between input and output of Lossy Compression**

Lossless compression technique is reversible in nature, whereas lossy technique is irreversible. This is due to the fact that the encoder of lossy compression consists of quantization block in its encoding procedure.

**JPEG**

JPEG (pronounced "jay-peg") is a standardized image compression mechanism. JPEG also stands for Joint Photographic Experts Group, the original name of the committee that wrote the standard. JPEG is designed for compressing full-color or gray-scale images of natural, real-world scenes. It works well on photographs, naturalistic artwork, and similar material. There are lossless image compression algorithms, but JPEG achieves much greater compression than with other lossless methods.

JPEG involves lossy compression through *quantization* that reduces the number of bits per sample or entirely discards some of the samples. As a result of this procedure, the data file becomes smaller at the expense of image quality. The usage of JPEG compression method is motivated because of following reasons:-

a. The *compression ratio* of lossless methods is not high enough for image and video compression.

b. JPEG uses *transform coding*, it is largely based on the following observations:

Observation 1: A large majority of useful image contents change relatively slowly across images, i.e., it is unusual for intensity values to alter up and down several times in a small area, for example, within an 8 x 8 image block.

Observation 2: Generally, lower spatial frequency components contain more information than the high frequency components which often correspond to less useful details and noises.

Thus, JPEG is designed to exploit known limitations of the human eye, notably the fact that small color changes are perceived less accurately than small changes in brightness. JPEG can vary the degree of lossiness by adjusting compression parameters. Also JPEG decoders can trade off decoding speed against image quality, by using fast but inaccurate approximations to the required calculations. Useful JPEG compression ratios are typically in the range of about 10:1 to 20:1. Because of the mentioned plus points, JPEG has become the practical standard for storing realistic still images using lossy compression.

JPEG (encoding) works as shown in the figure. The decoder works in the reverse direction. As quantization block is irreversible in nature, therefore it is not included in the decoding phase.

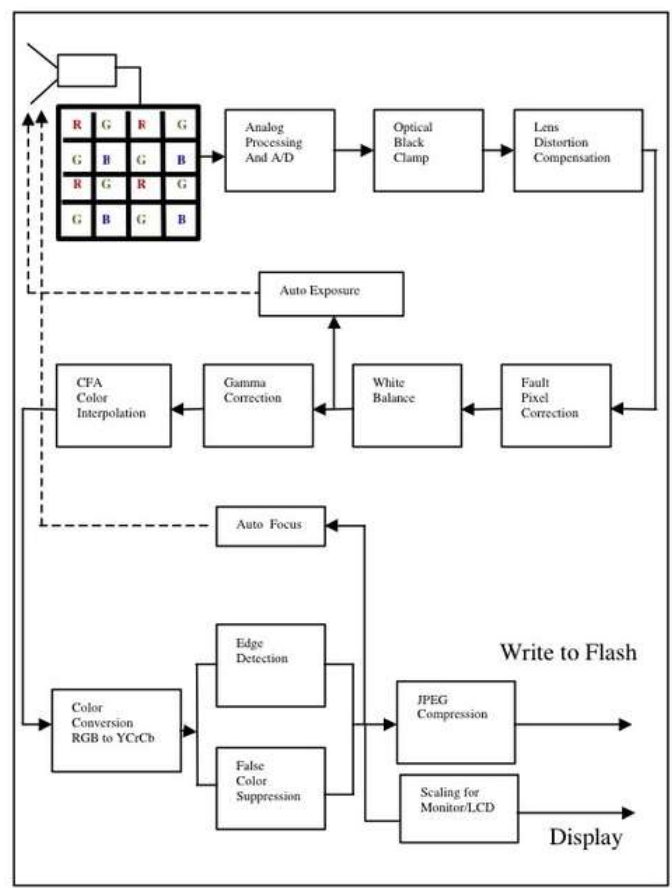

**Fig 5 Steps in JPEG Compression**

A major drawback of JPEG (DCT-based) is that blocky artifacts appear at low bit-rates in the decompressed images.

Such artifacts are demonstrated as artificial discontinuities between adjacent image blocks.

An image illustrating such blocky artifacts is shown in the figure 6 below:-

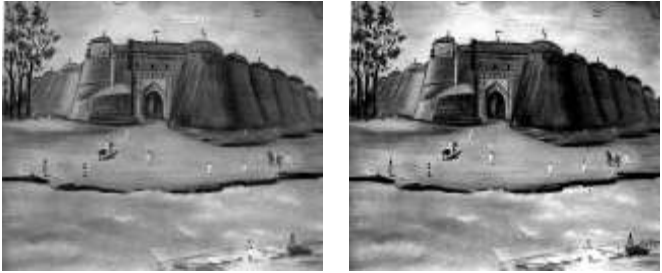

**Fig 6 : a).Actual Image      b).Blocked Image**

This degradation is a result of a coarse quantization of DCT coefficients of each image block without taking into account the inter-block correlations. The quantization of a single coefficient in a single block causes the reconstructed image to differ from the original image by an error image proportional to the associated basis function in that block.

Measures that require both the original image and the distorted image are called "full-reference" or "non-blind" methods, measures that do not require the original image are called "no-reference" or "blind" methods, and measures that require both the distorted image and partial information about the original image are called "reduced-reference" methods.

Image quality can be significantly improved by decreasing the blocking artifacts. Increasing the bandwidth or bit rate to obtain better quality images is often not possible or too costly. Several approaches to improve the quality of the degraded images have been proposed in the literature. Techniques, which do not require changes to existing standards, appear to offer the most practical solutions, and with the fast increase of available computing power, more sophisticated methods, can be implemented. The subject of this dissertation is to salvage some of the quality lost by image compression through the reduction of these blocking artifacts.

**CONCLUSION**

It is very much clear from the displayed results that the artifacts are removed to some extent as it has increased the subjective as well as objective quality of the images. The algorithm effectively reduces the visibility of blocking artifacts along with the preservation of edges. It also increases the PSNR value of the image. As shown, the blocking artifacts are not removed totally. It is because of the fact that the information lost in the quantization step is irrecoverable. The algorithm only deals with pixel values (spatial domain) and the algorithm only tries to manipulate the pixel values on the basis of some criteria. The extent of blocking artifacts can also be reduced by manipulating the DCT coefficients, as quantization is applied on the DCT coefficients. But frequency domain is having higher complexity and consumption of time. There is always a tradeoff between time (complexity) and efficiency (quality). Spatial domain is chosen where time (complexity) is the main concern, and on the other hand frequency domain is preferred where efficiency is given more value.The extent of reduction of blocking artifacts can be increased by recovering the information loss by using some sort of prediction algorithm. It can be done by some learning technique (artificial intelligence) or fuzzy logic.